# Understanding Twitter's behavior during the pandemic: FakeNews and Fear

Guillermo Romera Rodriguez, Sanjana Gautam, Andrea Tapia

Pennsylvania State University
{gkr5144, sqg5699, axh50}@psu.edu

**Abstract**

The outbreak of the SARS-CoV-2 novel coronavirus (COVID-19) has been accompanied by a large amount of misleading and false information about the virus, especially on social media. During the pandemic social media gained special interest as it went on to become an important medium of communication. This made the information being relayed on these platforms especially critical. In our work, we aim to explore the percentage of fake news being spread on Twitter as well as measure the sentiment of the public at the same time. We further study how the sentiment of fear is present among the public. In addition to that we compare the rate of spread of the virus per day with the rate of spread of fake news on Twitter. Our study is useful in establishing the role of Twitter, and social media, during a crisis, and more specifically during crisis management.

# 1 Introduction

In December 2019, a novel coronavirus, SARS-CoV-2, was identified as the pathogen causing coronavirus disease (COVID-19) in Wuhan, China. On March 11, 2020, COVID-19 was declared a pandemic by the World Health Organization (Muniyappa & Gubbi, 2020). The spread outside China was initially to other countries in Asia, most notably the Republic of Korea, then to the Middle East, most notably

Iran, then to southern Europe, most notably Italy and Spain, then further north in Europe, the UK and then to the most recent epicentre; the USA. The symptoms of the virus could show up anywhere from 2-14 days [1] (with a median of 5 days), or be completely asymptomatic, which caused the virus to spread like wildfire. Having spread throughout the world at a fast rate it was just a matter of time until its presence was greatly felt in the United States. As of February 2021, the United States has recorded a total of 26,523,297 cases [2]. In these early days this caused many healthcare systems across the world to be instantly overwhelmed and a high rate of mortality.

Within weeks of the emergence of the novel coronavirus disease 2019 (COVID-19) in China, misleading rumors and conspiracy theories about the origin circulated the globe paired with fearmongering, racism, and mass purchase of face masks, all closely linked to the new 'infomedia' ecosystems of the 21st century marked by social media (Depoux et al., 2020). COVID-19 spread quickly in part because of misinformation on social media platforms. It is therefore critical that public health responses manage not only the spread of infectious diseases but also the spread of misinformation on social media. Misinformation and incomplete information concerning the source of spreading may have prevented people from taking the right precautions. In initial cases, the lack of information about the severity of the situation may have led people to not observe caution (Lancet, 2020). During the initial months of the pandemic the information about its infection mechanisms were still under scrutiny (Adhikari et al., 2020) and even the use of personal protective equipment (PPE) such as masks, disinfectant, gloves wasn't entirely validated. Moreover, its lethality and spread was initially thought to be that of a simple flu by some sectors of the population which further impacted the spread of the virus.

In the absence of an effective treatment or vaccine, researchers pointed out that managing the pandemic response would require leveraging insights from the social and behavioral sciences, particularly with regard to non-pharmaceutical interventions and containing the spread of misinformation about COVID-19 (van der Linden, Roozenbeek, & Compton, 2020)(Habersaat et al., 2020)(Depoux et al., 2020).

According to Wilson and Chen (2020) (Wilson & Chen, 2020), social media panic spread faster than the COVID-19 outbreak. This can be supported by the first quarter (2020) report by Twitter that indicates that the average daily users had increased by 24% over the previous year to 166 million daily users, attributing much of that increase to activity resulting from the COVID-19 pandemic (Tsui, Rao, Carey, Feng, & Provencher, 2020). When users like, comment, or retweet a post, it also can increase the original post's impressions. Users on Twitter can amplify a post further by making a comment while retweeting another user's post. These new posts additionally generate their own set of impressions, which are not counted in the original post's impressions, thereby increasing visibility (Tsui et al., 2020). With the popularity of the internet and smartphones, information on COVID-19 spread rapidly among the public on social media. On social media, some people have shared false claims because they failed to think about whether the content was reliable (Radwan & Radwan, 2020).

[1]https://www.cdc.gov/coronavirus/2019-ncov/symptoms-testing/symptoms.html
[2]https://covid.cdc.gov/covid-data-tracker

In times like these, factual and truthful information is paramount. The major component for the positive influence of social media is concentrating on the exchange of corrected and balanced information and positioning of the public in a proper environment that presented the necessary health information and thus at the end enables the public to contain the COVID-19 pandemic (Radwan & Radwan, 2020). The spread of false information, incorrect facts, can generate panic, fear and anxiety in already troubled times. In addition to false information, we also need to consider the overall sentiment of the population, and how these measures impact them. Not only because of the overall wellbeing of the population but because it also impacts the abiding of these rules. Misinformation causes confusion and spreads fear, thereby hampering the response to the outbreak. The paper aims to inform the possible design interventions to help the management and spread of misinformation during a pandemic. Within this context our paper explores and examines the dialogue and the amount of fake news that were present and how they differed from real news during the month of august 2020.

# 2 Related Work

## 2.1 Fake News and Crisis

Fake news is a topic that it has gained traction and attention in the past few years with the emergence of modern machine learning algorithms and AI, and some argue the US presidential election of 2016 (Lazer et al., 2018; Gelfert, 2018; Lee, 2019). But what can be defined as fake news? To define fake news in the context of our study we need to look at the current literature and how fake news is defined in the domain of social media. A common definition for fake news can be found in the study by Allcott and Gentzkow (Allcott & Gentzkow, 2017), in where they describe fake news as ”news articles that are intentionally and verifiable false, and could mislead readers”. According to their definition we can see that satirical news sources could fall in that category of fake news. However, the latter differ in that their intent and organizational process are vastly different (Lazer et al., 2018). Fake news constitute an issue insofar it can create unwanted behaviors and opinions based on false information which can dangerous for society and communities alike (Narwal, 2018)

In the recent years we have seen a steady increase in the amount of fake news that have been created and propagated. People may spread fake news due to a variety of different reasons like attracting audience towards website to generate more revenue, to seek attention of people, to influence opinions of people and many more (Narwal, 2018). In our case, we feel the reasons might differ from the general ones. First, social media and access to those is more widespread than ever and its use as a source of information (Westerman, Spence, & Van Der Heide, 2014). Also, fake news are harder to detect by traditional means (Strickland, 2018). Moreover, these algorithms, sometimes called ”Bots”, can be highly automated without the need for any human supervision, which can allow for a high output of fake news in multiple domains and almost in real-time (Shao, Ciampaglia, Varol, Flammini, & Menczer, 2017). Lastly, we have the rise of fake news with the goal of destabilizing and reducing the credibility of trusted news platforms and swaying public opinion (Narwal, 2018). In addition, we have seen also an increase activity with the goal of destabilizing and undermining one's country source of news and spreading fake information with the intention of polarizing population's perspectives and ideas.

The COVID19 pandemic has been plagued with fake news and misinformation since its inception and spread at an alarming rate (Kouzy et al., 2020). Countries such as Italy, saw a sharp increase in fake news related COVID19 (Moscadelli et al., 2020). While some fake news and conspiracies were less damaging, some others carried more weight and deleterious consequences (Naeem, Bhatti, & Khan, 2020). For example, in countries such as Spain, the distrust on official news sources during the initial weeks of the pandemic led to an increase of social media as source of news (El´ıas & Catalan-Matamoros, 2020). In others, they are a marker for underlying and pre-existing beliefs (Georgiou, Delfabbro, & Balzan, 2020). In some other cases, fake news can pose a public health risk by giving the citizenry false or fake information that goes against scientists and health officials' recommendation and guidelines (Naeem et al., 2020).

To find the extent of fake news and disinformation strategies some studies found that as early as February there already were social media accounts (Twitter) using different hashtags to spread misinformation and information that could not be verified (Kouzy et al., 2020). It is why understanding the extent of the intrusion of fake news within the domain of social media is paramount to inform and and provide better approaches to solve it.

## 2.2 Use of Machine Learning Algorithms in Fake News Detection

There is evident success in detection of fake news and posts using various Machine learning approaches (Manzoor, Singla, et al., 2019). There has also been considerable work done in comparing various machine learning (ML) to identify the best performing ones. Among those, it has been seen that although a neural networks model has high computational requirements it outperforms other models in terms of accuracy. The stochastic gradient (SGD) classifier is the fastest learning algorithm. Support Vector Machine (SVM) uses the concept of hyperplanes instead of probability distributions, as seen in Naive Bayes. (Ahmad & Lokeshkumar, n.d.) SVM is the actual classifier and GD (or SGD) is the function that tells the classifier how correctly it has done the prediction based on which the SVM classifier improves itself. This establishes SVM as a sound model for our work. Furthermore, some works have presented a detection model for fake news using n-gram analysis through the lenses of different features extraction techniques. An investigation of two different feature extraction techniques and six different machine learning algorithms revealed that the model achieves its highest accuracy of 92% when using unigram features and Linear SVM classifier (Ahmed, Traore, & Saad, 2017). Among our three pre-training models, CountVectorizer achieves in general the best performance comparatively and Word2Vec performs relatively poor amongst the three models (Gravanis, Vakali, Diamantaras, & Karadais, 2019). We designed our code around the learnings from previous literature in the field.

## 2.3 Sentiment Analysis using Twitter Datasets

Twitter data can be analyzed in a myriad of ways (Ahmad & Lokeshkumar, n.d.), and depending on the task it can take different forms. Generally, Twitter data is mostly analyzed in its textual form. That is analyzing each tweet textual data to gather insights and find common themes. However, this poses some difficulties as the language used in Twitter can vary wildly from user to user, and Twitter's limitation of 280 characters. In addition to these difficulties, we have issues such as identifying sarcasm, metaphors and other types of languages that can be difficult identifying by the written word (Go, Huang, & Bhayani, 2009). However, one of the great benefits of social media data, in this case Twitter, is that it provides an almost real-time, and backwards looking window to events.

Even if we understand the benefits of using social media data to understand social discourse, crisis management, and overall sentiment, this does not make the task of using the data any easier (Kouloumpis, Wilson, & Moore, 2011). The approaches to identify fake news and misinformation are very varied. For example, there are some approaches that focus their efforts on working with fake news datasets (Wang, 2017; Buntain & Golbeck, 2017). These are crowdsourced datasets in which machine learning models and different algorithms can be tested and refined for later implementation. This approach has the advantage that the fake news and misinformation text has been previously labeled and verified. However, the issue with these approaches is the lack and variety of datasets in that not a single dataset can provide all features of interest (Shu, Sliva, Wang, Tang, & Liu, 2017). Other approaches focus their efforts on the pre-processing techniques that are applied to the text, leveraging any possible discrepancies that can be present between truthful and fake news (Davis & Proctor, 2017).

# 3 Research Context

As we have already established in our literature review, Twitter has played a pivotal role in case of information propagation during disasters in the past. Similarly, when it comes to a global pandemic it did have an impact on people. In our paper, we try to study the amount of fake news circulating on the social platform. We have seen over the course of this year that fake news, conspiracy theories (CT), disinformation, and false claims have roamed freely through social media. For example, the study by Budhwani and Sun (Budhwani & Sun, 2020) showed that the number of people mislabeling the COVID19 virus as the ”Chinese virus” increased substantially over the period of one month. Their paper serves as a testimony on how social media (Twitter in this case) has far-reaching implications and can perpetuate certain stigmas. The second component of our work revolves around the sentiments behind the Tweets. Depending on the sentiment captured on the Tweet, we can extrapolate the general reaction of the public. We can also identify the source of disruptions. This would help in the assessment of the far-reaching effect of fake news. We essentially are discussing the following two questions:

- *What proportion of Tweets being circulated consist of fake news related to Covid-19?*
- *How strongly the feeling of fear resonated among Twitter users as reflected by their Tweets on Covid-19?*

# 4 Twitter Data Processing

## 4.1 Twitter API Collection

Twitter at its core is a micro-blogging platform, where each user is able to post a short message, up to 280 characters, and view, follow, and like the activity created by other users. The platform, conceived as a means to track news by users everywhere (Kouloumpis et al., 2011), as gained notoriety in the last few years due to its difficulties to crack down, verify and fact-check disinformation tweets and fake news. This is problematic because Twitter user base can be counted in the hundreds of millions of active users monthly (Huberman, Romero, & Wu, 2008) that means fake news and disinformation can spread like wildfire and reach a variety of audiences. However, there are ongoing attempts to stop the spread of fake news and disinformation within the platform.

To collect information and attempt to study fake news and disinformation within the Twitter platform we rely on the different methods of collecting the data available in its platform and some external ones. Most systems that Twitter allow are through its proprietary API, and it allows users to collect data posted on the platform in almost real-time. In addition to its proprietary API, twitter allows access through other means, such as the Tweepy library (Roesslein, 2020) (Although API credentials are still required). This way allows researchers to collected data based on specific keywords, date and time, geographical location, and even language. The keywords used to retrieve the data can be crucial to look for specific themes and events. In our study the list of keywords can be seen in Table 1 (with some keywords removed for anonymization purposes). We decided to use these keywords as they were all referring in one way or the other with the ongoing pandemic. While we continuously collected data for the whole month of August, our efforts in this paper focus on the final weeks of August (from the $19^{th}$ through the $29^{th}$). The goal was to explore and investigate fake news in the pandemic context where the upward trend was not too extreme as to not bias our own results.

Table 1: Keywords used for data collection

| | |
|---|---|
| covid | covid19 |
| coronavirus | covidsars |
| covid19 | coronavirus |
| covidvirus | corona |
| outbreak | pandemic |
| communityspread | pandemia |
| rona | covid-19 |
| mask | ppe |
| flattenthecurve | stayathome |
| stayhome | socialdistance |
| sanitize | testing |
| lockdown | staysafe |
| quarantine | Maskup |

## 4.2 Pre-Processing Tweets

Social media data can be very difficult to work with, specifically due to its varied nature. The data can come in form of text, images, audio, and video. However, in our paper we are interested on the text data that we obtained through Twitter. Working with text data from a social media such as Twitter has its issues, and those are usually related to the ”cleanliness” of said text data. Therefore usually text data needs a certain level of pre-processing before it can be used for any purpose. As a rule the pre-processing state is composed of:

- Tokenization
- Noise removal
- Lower casing the text
- Stop-word removal (punctuation, hashtags, URLs, etc.)
- Theme-specific word removal

Arguably the most important step here is tokenization. Tokenization consists in separating individual characters (such as words) into smaller units, called Tokens. This is such an important step because unlike other social media, Twitter has a restriction of 280 characters. Because of this 280-character rule, users are constrained to fit as much information as possible in that given space. This generates a language that is composes of reduced words, slang, abbreviations, emoticons and more. Tokenizing words allow NLP and Machine Learning models to generate more effective and accurate results.

# 5 Methodology for fake news detection

To detect any possible instances of fake news and misinformation, we relied on previous methodologies and literature (Ahmed et al., 2017) Since our paper is focus on finding instances of fake news and misinformation, we did not focus our efforts on creating a custom made model but instead we did test multiple pre-built models, edited their hyper-parameters and compared their results to find the model that fitted our goal the best.

To optimize our model to the best of our knowledge and ability we did use three different datasets to train our models. Two out of the three datasets were collected from Kaggle [3]. Kaggle is a website that hosts, among many other things, data science competitions and it has been used in multitude of different domains and fields (Taieb & Hyndman, 2014; Mangal & Kumar, 2016; Iglovikov, Mushinskiy, & Osin, 2017; Yang et al., 2018). To select our datasets, we parted with the assumption that the COVID-19 pandemic was a disaster, and as such its behavior would be fairly similar to that encounter in other natural hazards and disasters. With this in mind we decided to use two datasets from Kaggle. The first dataset comes from a Kaggle competition titled *Natural Language Processing with Disaster Tweets* [4]. This dataset consists of a series of tweets related to disasters, and they can either be fake or not. The second dataset, very similar to our first one, titled *Disaster Tweets*[5] was created as an add-on to the first dataset we mentioned. The third dataset was not collected from a website but instead was a created by randomly concatenating the first and the second dataset. This was done with the intention of training and testing our models as much as we possibly could.

For all three datasets the testing and testing phase was the same but consisted of two distinct parts. This was done with the intention of optimizing computing time and power, since our resources were limited given the size of our original dataset. The first part of our work was focused on the pre-processing of the data. To do so we resorted to the package *ScikitLearn* and its function *TF-IDF vectorizer*. Like other functions such as the ones in the package *nltk*, this function cleans the data automatically, including noise removal, tokenization, lower casing, and stop-word removal. The only difference in this function is the use of a matrix of term frequency–inverse document frequency (TF-IDF), which is a statistical measure to better evaluate the number of times a word appears in a text with the improvement that words that tend to repeat many times within a single text *"the"* lose weight over others. After the cleaning process, we moved onto the second part, that is the training of our models and testing them. For our training and testing we used six different machine learning model across the three datasets. These being *Naive Bayes, k-nearest neighbor (KNN), Passive-Aggressive Classifier, Artificial Neural Network (ANN), XGBoost*, and *Linear Support Vector (SVM)*. The metrics we used to evaluate these models were accuracy, and F-score, also called the F1 score, which measures the models' accuracy based on the harmonic mean of the models' precision and recall, where 1 is the best possible score and 0 the worst. Once we ran the models and compared their results (Tables 2,3, and 4) we selected the model than performed the best from the outset in all three datasets. As seen in Tables 2,3, and 4 we can see that the model that performed best was SVM, which should not comeas a surprise after reviewing modern literature in this domain. Once we selected SVM as our preferred model we then tuned its hyper-parameters to optimize it during its training phase. We did so using the function *GridSearchCV* from the package *ScikitLearn*, which allowed us to get up to a *85%* accuracyand a F1 score of *0.7113*.

[3]https://www.kaggle.com/
[4]https://www.kaggle.com/c/nlp-getting-started/data?select=train.csv
[5]https://www.kaggle.com/vstepanenko/disaster-tweets

| **Models** | **Accuracy** | **F1 Score** |
|---|---|---|
| Naive Bayes | 0.79382 | 0.72600 |
| KNN | 0.78594 | 0.73581 |
| PAC | 0.78069 | 0.75185 |
| ANN | 0.77544 | 0.74169 |
| XGBoost | 0.73473 | 0.63003 |
| SVM | 0.80499 | 0.77276 |

Table 2: Model Results with first dataset

| **Models** | **Accuracy** | **F1 Score** |
|---|---|---|
| Naive Bayes | 0.83773 | 0.25454 |
| KNN | 0.87598 | 0.624 |
| PAC | 0.87554 | 0.67283 |
| ANN | 0.87467 | 0.66821 |
| XGBoost | 0.84740 | 0.34896 |
| SVM | 0.89226 | 0.69104 |

Table 3: Model Results with second dataset

| **Models** | **Accuracy** | **F1 Score** |
|---|---|---|
| Naive Bayes | 0.81195 | 0.52336 |
| KNN | 0.82170 | 0.64198 |
| PAC | 0.82038 | 0.68686 |
| ANN | 0.8230 | 0.68685 |
| XGBoost | 0.79220 | 0.49904 |
| SVM | 0.84856 | 0.71964 |

Table 4: Model Results with third dataset

# 6 Methodology for sentiment analysis

Next for analyzing the sentiments, we followed a similar process to clean the twitter data. We used the text2emotion library in Python. Text2emotion[6] is the python package that helps in extracting the emotions from the content. It processes any textual message and recognize the emotions embedded in it. Compatible with 5 different emotion categories as Happy, Angry, Sad, Surprise and Fear. We used this package for its ability to extract 'Fear' instead of positive/negative emotions that are obtained from sentiment anaylsis techniques. The way the package works is, at first the major goal is to perform data cleaning and make the content suitable for emotion analysis. Detect emotion from every word that we got from pre-processed text and take a count of it for further analytical process. Find the appropriate words that express emotions or feelings. Check the emotion category of each word. Store the count of emotions relevant to the words found. The output will be in the form of dictionary. There will be keys as emotion categories and values as emotion score. Higher the score of a particular emotion category, is conclusive of the fact that the message belongs to that category. This information is relayed to the user by means of a python dictionary with 5 keys referring to the 5 emotions that can be analyzed by this package; the value of these keys is the proportion that a given emotion is present in the text (Ramírez-Sáyago, n.d.). Notably, the sum of the values of the emotions must be either 1 or 0, the latter only appearing in cases where no emotion could be detected in the text (such as for nonsensical strings of characters). For example, a piece of text that is exclusively angry would have an'angry' value of 1, and every other emotion would have a value of 0. On the other hand, a message that is in equal parts happy and surprised, would have a 'happy' value of 0.5, a 'surprise' value of 0.5, and every other emotion would have a value of 0. 2 different sets of analysis were performed in this research, the first of these was done to the set of tweets.

[6]https://shivamsharma26.github.io/text2emotion/index.html

# 7 Fake news results

After having trained and tested our SVM model we then applied it to our dataset consisting of 3,107,71. According to our model 2,673,410 tweets are classified as fake news or false information, and only 434,306 real news. This accounts for 86% of all the tweets we collected during the month of august to be fake news or false information. Yet, we cannot assume that all those tweets were fake news, especially when our best performing model had an accuracy of 80.7%. However, that is why we attempted to adhere to previous literature and follow a similar methodology for training and testing the models, as well as pre-processing and cleaning the data. We also are aware of the difficulties that arise from analyzing social media data such as tweets. To give a more in-depth detail we compared the N-grams of the fake news subset of data with the real news. The results as shown in Table 5, tell a story about the difference in language that both subsets of data had.

| **Fake News (Top 3 N-grams(N=3,4,5))** | **Real News (Top 3 N-grams(N=3,4,5))** |
|---|---|
| (nasdaqpill, form, vxrt, manufacture, oral) | (sunglass, give, level, anonymity, desired) |
| (kin, nasdaqpill, form, vxrt, manufacture) | (mask, sunglass, give, level, anonymity |
| (manufacture, oral, vaccine, candidate, covid19) | (wearing, mask, sunglass, give, level) |
| (manufacture, oral, vaccine, candidate) | (wearing, mask, sunglass, give) |
| (vxrt, manufacture, oral, vaccine) | (sunglass, give, level, anonymity) |
| (oral, vaccine, candidate, covid19) | (give, level, anonymity, desired) |
| (vaccine, candidate, covid19) | (american, died, covid19) |
| (manufacture, oral, vaccine) | (wearing, mask, sunglass) |
| (vxrt, manufacture, oral) | (tested, positive, covid19) |

Table 5: Comparison of Fake and Real N-grams

From the table we can see that the first thing we noticed was the difference, not only in the words used, but also the theme that it seemed to carry in it. The fake news subset of data had its conversation revolving around the manufacturing, production, process of the vaccine, and even the companies involved in it. This would coincide with the timeline in which the first vaccines were showing results and the process was some months away from completion. Overall, the words used on the fake news subset seem to suggest a more speculative language in regards to the COVID-19 vaccine. On theother hand, the subset of real news displayed a noticeable difference not only in the words but also onthe theme that was mentioned. The words and the theme seemed to be more in line with the overall pandemic behavior. This type of behavior might not be too surprising as vaccines have been an objectof misinformation and false information during the last few years (Chiou & Tucker, 2018), and the COVID-19 pandemic has not been different as shown in Table 5.

# 8 Sentiment Analysis Results

After processing 3 million Tweets, we finally had our results on people's sentiments. As we mentioned before, our results are represented as a percentage. We had the following results for different sentiments:

happy (10%), sad (27.5%), angry (5%), surprise (21.3%) and fear (36.3%). This can be seen in Figure 1. As you can see the highest representation is for sadness and fear. This is not at all surprising keeping in mind that during that period COVID19 had taken a turn for worse in the United States of America. A thing to note about the results is that this analysis was done on over 3 million Tweets. In addition to that, we also ran a search for the top 50 terms used in the Tweets in Figure 3. As we can see the most used terms were ’covid’ and ’mask’. This is not at all surprising and quite expected. There was considerable dialogue surrounding wearing masks and following other preventive measures. This debate may have been the reason for the words that followed. Words like ’Trump’, ’testing’ and ’vaccine’ are indicative of the type of misinformation that may be spreading during the time. They represent certain politic air to the discussion that happened. It is natural to expect it was surrounding the topics discussed above. We also used a word cloud (figure 2) to represent the terms. The word cloud goes on to display other terms that were popular at the time. The word ’lockdown’ catches one’s attention.There was a lot of discussion surrounding both extensions and leniency of the lockdown. The words together with the sentiment paint a picture of the general climate of opinions. Fear surrounding the uncertainty of length of lockdown, effectiveness and arrival of vaccine, debate regarding mask wearingare some of the examples that come to mind.

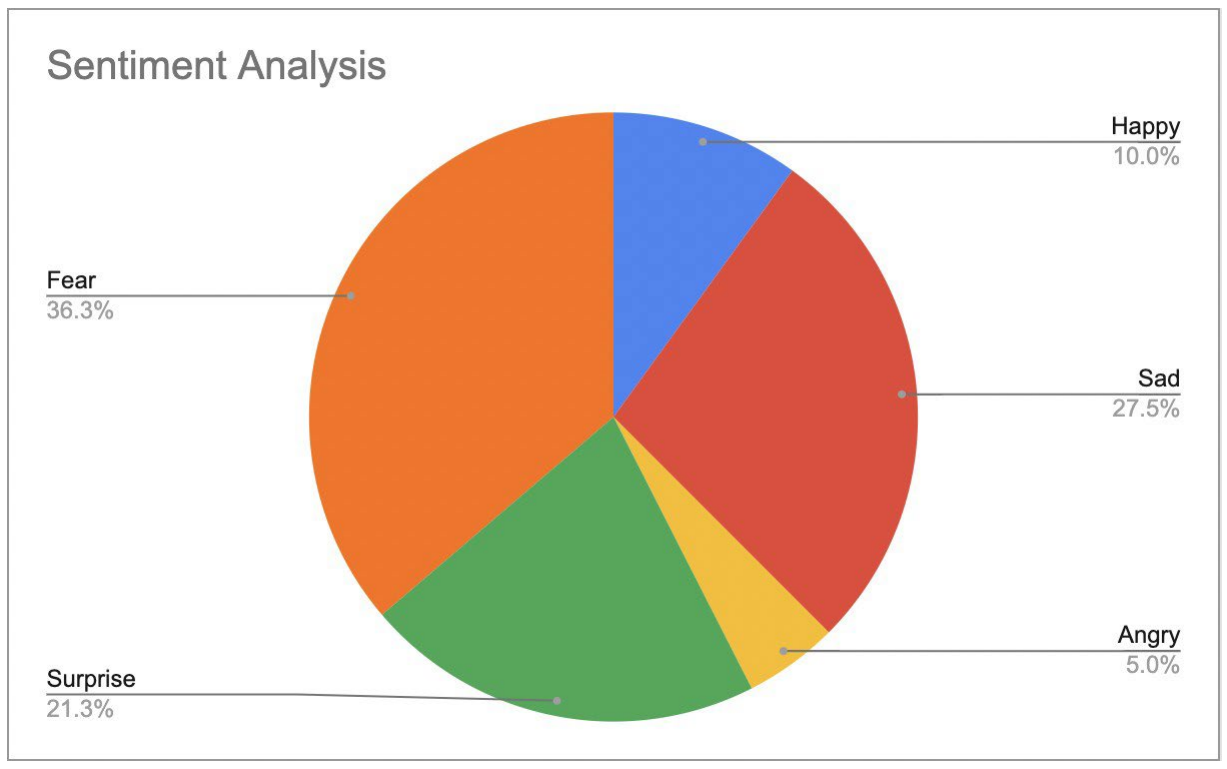

Figure 1: Twitter Sentiment Analysis on Fake News subset

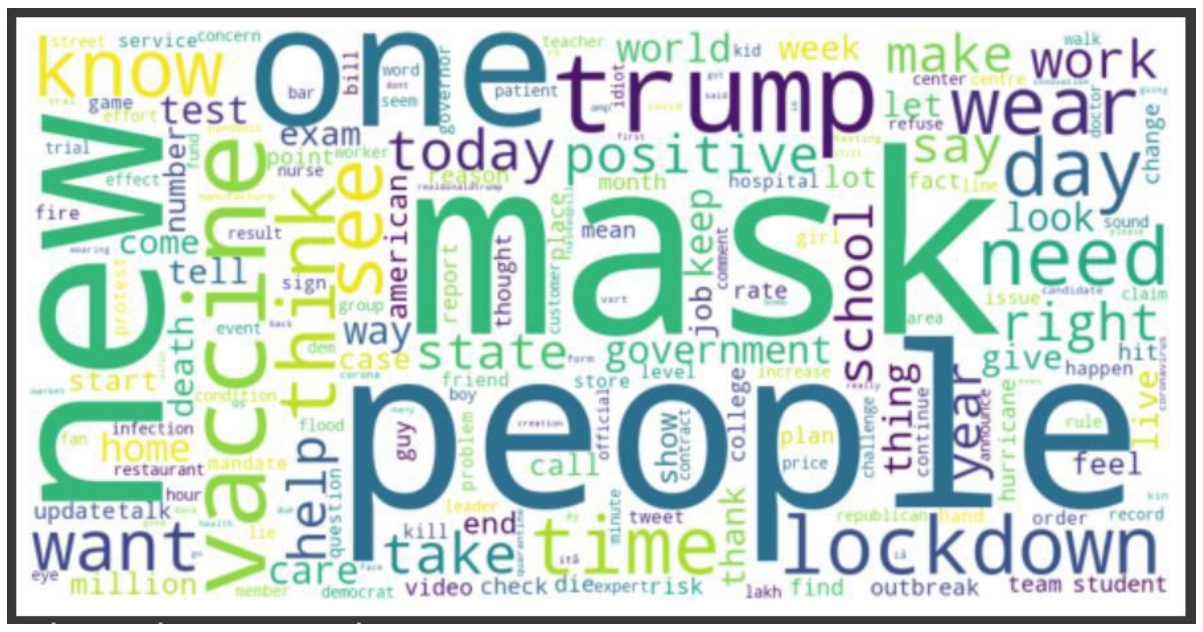

Figure 2: Daily count of fake news

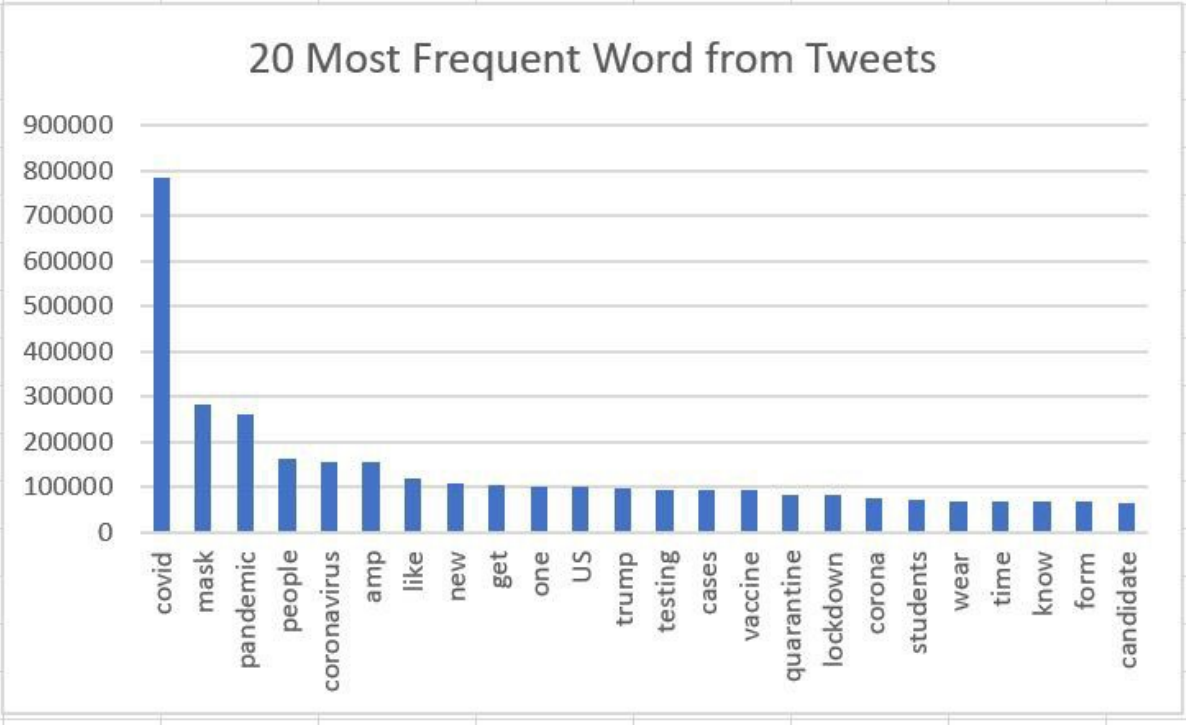

Figure 3: Daily count of COVID19 cases

# 9 Discussion

The work presented here is indicative of some important conclusions. Firstly, the spread of false beliefs is not merely a side effect of fake news but is a direct result of its function (Gelfert, 2018). Here we presented the results for all tweets that were collected for the month of August written in English. While we could not assume that all tweets were originated from a particular location, we know that certain segments of the US population were highly vocal about masks regulations during the early months of the pandemic (Scerri & Grech, 2020). This is worrisome insofar that adherence to these regulations is contingent to the information that people received and the outlook of making a change. This way certain demographics of the population caught by fake news and false information could lead to opposition of such regulations. Promoting accurate beliefs about COVID-19, and encouraging healthier, safer behaviors related to COVID-19 prevention, would certainly answer this call (van der Linden et al., 2020).

The sentiment analysis showed the skewness of emotions found within our data, and on fake news. The most prevalent emotion was that of *fear* followed by *Sad* and *Surprise*. It could be the case that these fake news and instances of disinformation would be done to cause hysteria, and widespread panic. This can be certainly perilous in the case of fake news and disinformation as this panic can lead to unwarranted and unwanted behaviors from the general population. Nonetheless, this is an example of the importance of managing and using real information and be on the lookout for fake news and false information during period of public crisis, as it can lead to a more widespread panic, uncertainty and dangerous behavior.

During the past years we have seen that social media can affect different aspects of social interaction and behavior. The COVID-19 pandemic has certainly changed the daily lives of millions of individuals. As such, social media has been one of the few methods of communication for some individuals as restrictions and lock-downs went into effect in different parts across the world. For many, social media was also their source of news. Therefore it is important to investigate the behavior shown in social media in events such as the COVID-19 pandemic. This study provides a window to the development the fake news spread during the pandemic. More specifically, studying and learning how social media and fake news played a role in the development of this crisis can help alleviate or create new frameworks that can be used in future situations. Building up on which, studying the sentiment response of the Tweets puts into perspective the impact of spread of fake news. Feelings of fear among the Twitter users captured is indicative of disruption caused in the time of pandemic by the propagation of fake news.

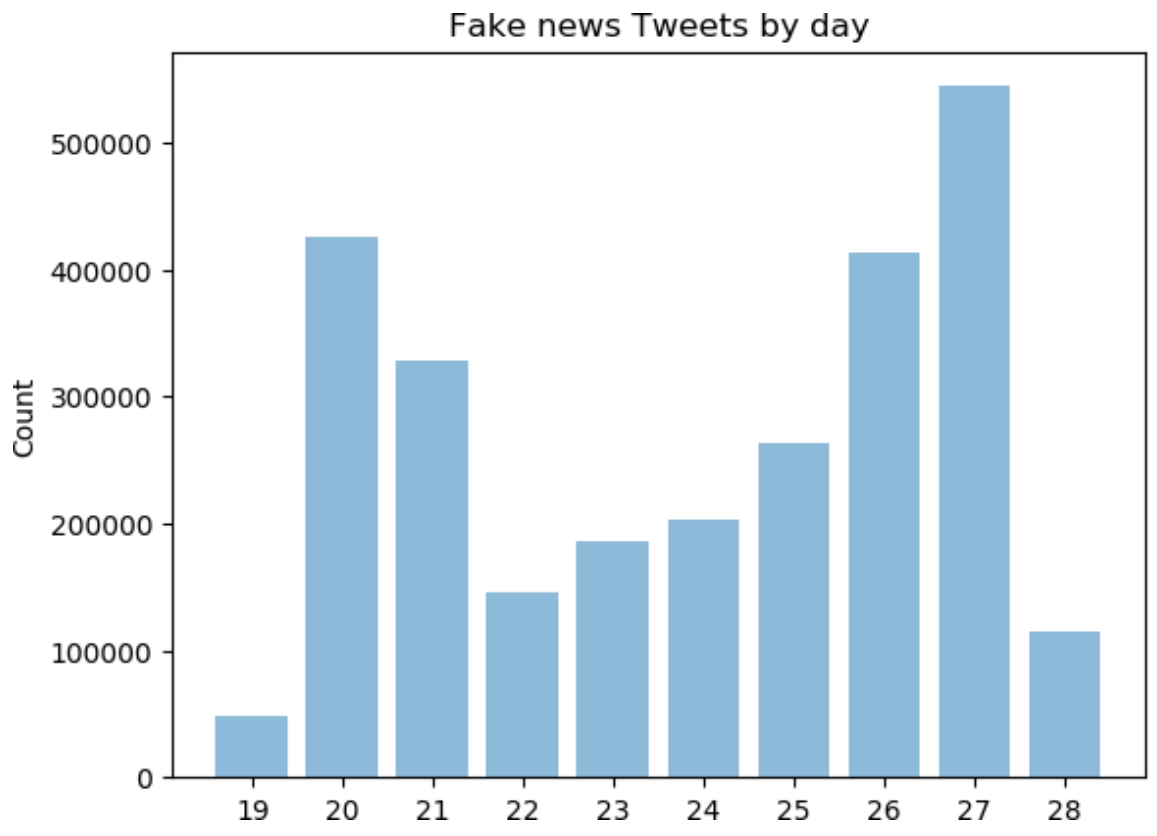

Figure 4: Daily count of fake news

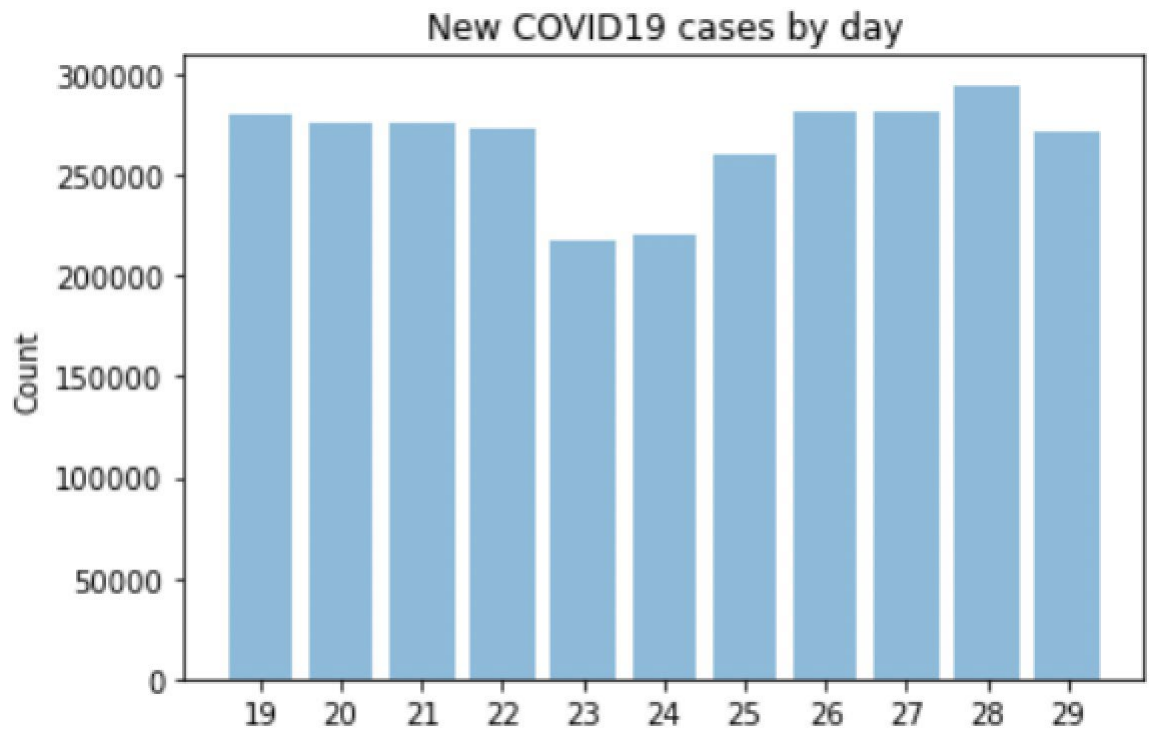

Figure 5: Daily count of COVID19 cases

Our study also shows the daily increase in cases with the recorded number of fake news spread during those days. While there is not a clear similarity, in Figure 4 and 5 it can be appreciated that there was a slight decrease both in fake news and covid cases that followed a rapid increase. It could be argued that this increased activity in social media could be affected by the numbers of positive cases. Further, as evident from the word cloud and most frequent words, ’mask’ and ’trump’ were popular keywords in the tweets captured. There was considerable debate on-going about adhering to wearing masks. This has been reflected in our analysis as well (Figure 3). Words like ’covid’ and ’pandemic’ were naturally expected to have been the most popular choice.

## 10 Limitations

As we mentioned before this analysis is done on the data that was collected over the month of August. We were faced with many challenges in the process. The processing of these many tweets took substantial time. All the data that was captured was in English, so the population being represented is English speaking. So, we can expect a bias. At this point it would also be useful to mention that detecting fake news using Tweets is challenging on account of it’s poor performance on sarcastic text (Liebrecht, Kunneman, & van Den Bosch, 2013).

## 11 Conclusion

In conclusion, we would like to say that our work offers for great insights into the public opinions during the COVID19 pandemic. Assessment of the public sentiment as well as accessible information is of great importance in the given scenario. All the measures adopted by the Governments to contain the virus heavily depended on the ability of the people to be aware and act responsibly. A collective response was the only way to ensure that the virus is contained and normal achieved. Thus, our work sets the groundwork to indicate the area in which control and work is required for better crisis response. The results from our study are indicative of the proportions of damage false public image of policies (in our case CDC guidelines) can cause. Further work is required to determine what measures may be adopted to curtail the damage done by the same. For now, we can say with certainty that social media platforms play a pivotal role in case of emergencies and their influence can be felt from the assessment of the results from our study.